\input harvmac.tex

\Title{}{Playing with  Neutrino Masses}\medskip

\centerline{Sheldon Lee Glashow\footnote{$^\dagger$}
{e-mail: slg@bu.edu}}
\bigskip\centerline{Physics Department}
\centerline{Boston University}\centerline{Boston, MA 02215}

\vskip .3in 

Most of what is known about neutrino masses and mixings results from
studies of oscillation phenomena. We focus on those neutrino
properties that are not amenable to such studies: $\Sigma\,$, the sum
of the absolute values of the neutrino masses; $m_\beta\,$, the
effective mass of the electron neutrino; and $m_{\beta\beta}\,$, the
parameter governing neutrinoless double beta decay. Each of these is
the subject of ongoing experimental or observational studies. Here we
deduce constraints on these observables resulting from any one of six
{\it ad hoc\/} hypotheses that involve the three complex mass
parameters $m_i$: (1) Their product or (2) sum vanishes;
(3) Their absolute values, like those of charged leptons or quarks of
either charge, do not form a triangle;  (4) The $e$-$e$ entry of the
neutrino mass matrix vanishes; (5) Both the $\mu$-$\mu$ and
$\tau$-$\tau$ entries vanish;  (6) All three diagonal entries are
equal in magnitude.  The title of this note reflects the lack of
any  theoretical basis for any of these simple assertions.

\Date{12/09}

\nref\rFi{G.L. Fogli {\it et al.,} Nucl.Phys.Proc.Suppl.188:27-30,2009.}
\nref\rFii{G.L. Fogli {\it et al.,} Phys.Rev.D78:033010,2008.}
\nref\rhz{Xiao-Gang He and A. Zee, Phys.Rev.D68:037302,2003.}
\nref\rsg{ P.H. Frampton, S.L. Glashow, D. Marfatia, 
Phys.Lett.B536:79-82,2002\semi
S.L. Glashow, {\tt arXiv: 0710.3719}}
\nref\rzee{A. Zee, Phys.Lett.B93:389,1980}
\nref\rslg{{\it E.g.,} A.Yu. Smirnov, Phys.Rev.D55:1665,1997\semi
P.H. Frampton and S.L Glashow, Phys.Lett.B461:95-98,1999\semi
Y. Koida, Nucl.Phys.Proc.Suppl.111:294,2002\semi 
Xiao-Gang He, Eur.Phys.J.C34:371,2004.}

\noindent
%{\it When in danger, when in doubt, run in circles, scream and
%shout!} \ ...Laurence J. Peter 
{\it And thus do we by indirection find directions out.}\quad Wm. 
Shakespeare\hfill
\medskip 

Much has been learned about neutrino masses and their mixing
 parameters. While numerous theoretical models and symmetry schemes
 have been proposed, none  evoke the clear ring of truth. The
 structure of the neutrino mass matrix remains largely unknown.
 Rather than launch yet another speculative model, here we explore a
 wide variety of potential simplicities.  \medskip

 We presume neutrino phenomena to involve just three left-handed
 states ($\nu_e,\,\nu_\mu,\, \nu_\tau$) whose Majorana masses are
 described in a flavor basis by the complex symmetric $3\times 3$
 matrix $\cal M$. We are not now concerned with the underlying origin
 of neutrino masses, only with their observable attributes.  Recall
 that the neutrino mass matrix may be written as:

$$ {\cal M}= U^*{\cal D}U^\dagger\,,$$
where $U$, the neutrino analog to the CKM  matrix,
is expressed in terms of the CP-violating parameter $\delta$ of the
neutrino sector and the
three mixing angles: 
$\theta_1\equiv \theta_{23}$, $\theta_2\equiv \theta_{31}$
and $\theta_3\equiv \theta_{12}$.  The diagonal
matrix $\cal D$ displays three complex mass parameters
 $m_i$, of which $m_3$ is
chosen to be real and positive with no loss of generality.     
\medskip

First  I  summarize what is known about these
parameters from current neutrino oscillation experiments\rFi, 
whose  best-fit  values will suffice for our subsequent  analysis:
$$\eqalign{\Delta_a &\equiv m_3^2 - (m_1^2+m_2^2)/2\simeq \pm 2390\;{\rm
(meV)^2}\cr
\Delta_s&\equiv m_2^2-m_1^2 \simeq 77\; \rm (meV)^2\cr
\cos{2\theta_1}\simeq 0.068 \quad\ \ &\sin^2{\theta_2}\le 0.030\quad\ \
\sin^2{\theta_3}\simeq 0.312\cr}$$
where we have adopted  a somewhat conservative upper bound for $\theta_2$.
\medskip

We  examine  the phenomenological consequences of any one of several
simple constraints that can be imposed upon $\cal M$.
Each of these constraints yields a neutrino mass matrix that is
fully compatible with all current oscillation data. 
Hence, we  study the effects of these constraints
on those   observable quantities that are not directly related
to such data, notably:

\vfill\eject

 (1) $\Sigma\equiv |m_1|+|m_2|+ m_3$, the sum of the 
 neutrino masses, for which
 several astrophysical upper bounds have been alleged\rFii,
 ranging  from 190~meV to 1190~meV.  For purposes of
our subsequent analysis, we arbitrarily adopt the bound $\Sigma\le 750\,$meV.
\medskip

 (2) $m_\beta\equiv \big(c_2^2c_3^2\,m_1^2+ c_2^2s_3^2\,m_2^2+s_2^2
\,m_3^2\big)^{1/2}\approx \big(c_3^2\,m_1^2+ s_3^2\,m_2^2\big)^{1/2}$,
 the rms mass of the electron neutrino,
which can reveal itself by precision
studies of beta-decay spectra, especially that of tritium.
We use the notation $s_3\equiv \sin{\theta_3}\,,\  m_1^2\equiv
|m_1^2|$,  \&c. Under most circumstances $m_\beta \simeq |m_1|\simeq |m_2|$. 
\medskip

 (3) $m_{\beta\beta}\equiv
\big|c_2^2c_3^2\,m_1+c_2^2s_3^2\,m_2+s_2^2\,m_3\big|\approx
\big|c_3^2\,m_1+s_3^2\,m_2\big|$, which  governs the
rate of neutrinoless double beta decay. Its value depends  on 
the relative  phase of $m_1$ and $m_2$, whose cosine we denote by $\mu$.
Under most circumstances (when $|m_1|\simeq |m_2|$) we have
 $m_{\beta\beta}\simeq m_\beta\sqrt{1-\zeta}$,
where $\zeta\equiv \sin^2{2\theta_3}\,(1-\mu)/2 \simeq 0.43\times(1-\mu)$.
The parameters $\mu$ and $\zeta$  play further  roles in what follows.

\medskip

We begin by  exhibiting   the bounds on these observables
resulting  from our  assumed upper limit to  $\Sigma$ and the known values of
$\Delta_a$ and $\Delta_s$, with no further hypotheses. 
Hereafter,  neutrino masses are expressed in milli-electron-volts:
$$\eqalign{
{\rm Normal\ Hierarchy:}\qquad & 750\ge \Sigma\ge 58\,,\quad
250\ge m_\beta\ge 5\,,\quad 250\ge m_{\beta\beta}\ge 0\cr
{\rm Inverted\ Hierarchy:}\qquad & 750\ge \Sigma\ge 98\,,\quad
375\ge m_\beta\ge 49\,,\quad 375\ge m_{\beta\beta}\ge 31}$$
%x
Current and anticipated measurements 
of  $m_\beta$ and $m_{\beta\beta}$ are summarized
elsewhere, {\it e.g,} \rFi.
We proceed to
 examine the consequences of any one of an  astonishing variety
of  additional hypotheses.
The results are  essentially unaffected by our  neglect
of terms involving  $s_2^2$ and,
 when appropriate, of those  involving
$\Delta_s$.

\medskip

{\bf I.} \  Suppose that $\det {\cal M}=0$. There are two relevant cases with
 one vanishing neutrino mass.
 (a) If $m_1=0$, the hierarchy is normal and:
$$\Sigma\simeq 57\;,\qquad m_\beta= 5\,,\qquad m_{\beta\beta} \simeq 3\,. $$
\indent (b) If $m_3=0$,  the  hierarchy is inverted and: 
$$\Sigma\simeq 98\,,\quad\  m_\beta \simeq 49\,,
\quad\  49\ge m_{\beta\beta}\ge 31\,.$$
\vfill\eject

{\bf II.} \ ${\cal M}_{ee}=0$, for which $m_{\beta\beta}$ is
set equal to zero {\it ab initio\/} and we obtain the relation:
$c_3^2\,m_1+ s_3^2\,m_2=0\,.$
The neutrino mass hierarchy is   normal and
all three observables are tiny:
$$\Sigma\simeq 64\,, \quad m_\beta \simeq 7\,, 
\quad m_{\beta\beta}=0\,.$$
\medskip

{\bf III.} \ Neither the masses of the three charged leptons, nor
those
of the three $Q={2\over 3}$ quarks, nor  those of the three
$Q=-{1\over 3}$ quarks  satisfy triangle equalities. Perhaps
that property is shared  by the three neutrino masses.
There are two cases.

(a) If the 
 hierarchy is normal, we require $|m_3|>|m_1|+|m_2|$.
Neglecting  $\Delta_s$, we
set  $|m_3|>
2|m_1|\simeq 2|m_2|$.
From $\Delta_a=|m_3|^2-|m_1|^2 $ we find:
 $|m_1|< \sqrt{\Delta_a/3}\simeq 29$ and $|m_3|< 2\sqrt{\Delta_a/3}\,,$
which imply:
$$58<\Sigma<116\,,\quad m_{\beta\beta}\le m_\beta < 29\,.$$
\indent  (b) If the hierarchy is inverted, we require
$|m_3|<|m_2|-|m_1|$. This may only be achieved for $|m_3|< 1$,
which effectively yields case Ib.
\medskip

\noindent The preceding scenarios would not please those 
experimenters seeking 
definitive measurements of $\Sigma$, $m_\beta$ or $m_{\beta\beta}$. Let
them read on!\medskip

{\bf IV.} \ Suppose $\cal M$ to be traceless so that $m_1+m_2+m_3=0.$
This possibility was first considered by He and Zee\rhz. We must
stress that the phases of the flavor eigen-fields may be adjusted
arbitrarily in the standard model or its simplest extensions.  Thus
the condition ${\rm Tr}{\cal M}=0$ is ordinarily senseless because it
is not not preserved by such phase redefinitions. Nevertheless one
may concieve of new physics that would fix these phases, thus rendering
the condition meaningful and making  its
consequences worth considering.
  Furthermore, we note (as did He and Zee) that in
general ${\rm Tr{\cal M}}\ne \Sigma\,m_i$.  However, their difference
is $\sim\!\sin^2{\theta_2}$, which we may safely ignore.  Our results
depend sensitively on the relative phase of $m_1$ and $m_2$, which
(due to the preceding argument) may not coincide with its value in the
Kobayashi-Maskawa ansatz.  We obtain:
$$|m_1| \simeq |m_2|\simeq \sqrt{\left|{\Delta_a\over 1+2\mu}\right|}
\qquad\quad m_3\simeq \sqrt{\left|{2(1+\mu)\,\Delta_a\over 1+2\mu}
\right|}\,. $$
Notice that neutrino masses approach degeneracy and
diverge as $\mu\rightarrow -{1\over 2}$.
The hierarchy is normal   for $\mu > -{1\over 2}$, inverted
for $\mu<-{1\over2}$, and nearly degenerate for  $\mu\sim -{1\over 2}$.
Several examples  may be worth consideration:
$$\eqalign{
{\rm (a)}\qquad   \mu &=1\qquad\ \ \  \Sigma=116,\quad
m_\beta=29,\quad\ \, 
m_{\beta\beta}=29\cr
{\rm (b)}\qquad  \mu &\simeq  -.5 \qquad \Sigma=750,\quad
m_\beta=250,\quad m_{\beta\beta}= 149\cr
{\rm (c)}\qquad   \mu &= -1\qquad\ \Sigma=98,\quad
m_\beta=49,\quad\  m_{\beta\beta}=31 \cr}$$
In (a), we have a normal hierarchy with $|m_3|\simeq 2|m_{1,2}|$.
In  (b), the neutrino masses are nearly degenerate and
$\mu$ is chosen to be  
close enough to $-{1\over2}$ for  $\Sigma$ 
to attain its  upper bound of 750~meV.  Case (c) yields a special case of Ib: 
an inverted hierarchy with $|m_3|=0$.\medskip

{\bf V.} \ Here we consider a previously studied\rsg\ model
wherein two diagonal entries of the neutrino mass matrix  vanish: 
 ${\cal M}_{\mu\mu}= {\cal M}_{\tau\tau}=0$.  
Two useful relations relations result from these constraints: 
$$\eqalign{
\cot{2\theta_1}\,\cot{2\theta_3} &= s_2\cos{\delta}\cr
\sin^2{\theta_3}\,m_1+ \cos^2{\theta_3}\,m_2 + m_3&=0\,.\cr}$$
The first relation correlates the small observed value of $\theta_2$
with the observation of maximal or nearly maximal atmospheric
oscillations. The second relation is relevant to the issues at hand,
from which we obtain:
$$ 
|m_1|= \sqrt{\Delta_a/\zeta}\,,\quad  \quad |m_3|=
\sqrt{\Delta_a(1-\zeta)/\zeta}\,,$$
where $\zeta$ was defined earlier. We examine three  examples:
$$\eqalign{
1-\mu= 2\ \ \ \; \qquad  &\Sigma=126\,,\qquad   m_\beta=53\,,\ \qquad
m_{\beta\beta}=32\cr
1-\mu=1\ \ \ \; \qquad &\Sigma=206\,,\qquad m_\beta=75\,,\ \qquad
m_{\beta\beta}=56\cr
1-\mu\simeq  0.09  \qquad &\Sigma\simeq  750,\,\qquad m_\beta\simeq
251\,, \qquad 
 m_{\beta\beta}=248\cr}$$
For  the third example we
 chose $\mu\simeq 0.91$,   small enough to saturate the bound on $\Sigma$.
The hierarchy is necessarily inverted and the 
neutrino masses both diverge and approach degeneracy as
$\mu\rightarrow -1$.
The relative (Majorana) phase of $m_1$ and $m_2$  is 
 related to the CP-violating phase $\delta$:
$$1-\mu = {2\cos{\delta}\over
\cos^2{\theta_3}+\cos{\delta}\sin^2{\theta_3}}$$
This relation shows  that $1\ge \cos{\delta}>0$, with a lower
bound of about 0.03 corresponding to our constraint on $\mu$.
Relatively large
 values of $m_\beta$ and $m_{\beta\beta}$ are permitted, 
but only for  $\delta$ 
close to  $\pi/2$, {\it i.e.,} for 
nearly maximal CP violation in oscillation phenomena.

\medskip {\bf VI.} \ Suppose  the three diagonal entries
of the neutrino mass matrix to be  equal in magnitude: $|{\cal
M}_{ee}|=|{\cal M}_{\mu\mu}|=|{\cal M}_{\tau\tau}|$.  With appropriate
phase redefinitions, this condition is seen to be equivalent to the
strictly off-diagonal 
Zee matrix\rzee, augmented by  a multiple
of the unit matrix. While the Zee model is experimentally
disfavored\rslg, this model is compatible with all available 
oscillation data.
We examine its  consequences  neglecting
relatively small terms involving $\Delta_s$,   $s_2$ and
$\cos{2\theta_1}$.
 With these  approximations, 
which cannot substantially alter our result,
the
condition reduces to a single real quadratic equation:
$$
R^2+2R(\mu_1\,c_3^2+\mu_2\,s_3^2)-3(1-\zeta)\,,$$    
where $R=|m_3/m_1|\simeq |m_3/m_2|$ and $\mu_i=\cos{\phi_i}$ with
$\phi_i$ being the Majorana phases of $m_1$ and $m_2$. Recall that we
have chosen $m_3$ to be real and positive and $\zeta=
\sin^2{2\theta_3}(1-\mu)/2$, where $\mu=\cos{(\phi_2-\phi_1)}$.
Again I offer three examples:\medskip

(a) $R$ attains its maximum value of 3
 for $\phi_1=\phi_2=\pi$.  This yields
a normal hierarchy with:
$$ \Sigma=190\,,\qquad m_\beta=m_{\beta\beta}=17\,.$$

(b) 
Degenerate neutrino masses result from the choice
 $\phi_1=\phi_2=0$, we obtain $R=1$.
Saturating our bound on $\Sigma$ near this 
point yields:
$$\Sigma=750\,,\qquad m_\beta=m_{\beta\beta}=250\,.$$

(c) 
The smallest possible value of $R$  is obtained for 
$\phi_1=-\phi_2=\pi/2$, whence $R\simeq 0.65$. This yields an inverted
hierarchy with:
$$\Sigma= 199\,,\qquad m_\beta=75\,,\qquad m_{\beta\beta}=57\,.$$

\medskip

Perhaps our discussion will prove stimulating to both experimenters
 and model builders. That so many disparate hypotheses remain
 compatible with what we already know about neutrinos reflects the
 depth of our ignorance about the  mechanism responsible for
 neutrino masses and mixings.  As experimenters and astrophysicists
 become ever more clever in  constraining  $\Sigma$,
 $m_\beta$ and $m_{\beta\beta}$, perhaps we shall  learn
 which, if any,  nature has chosen of   our six scenarios.
 
\bigskip
\centerline{ACKNOWLEDGEMENT}
I am grateful for  stimulating discussions with
Andrew Cohen. This research was  supported in part by the
 Department of Energy under grant number:
 DEFG02-91ER40676.
\listrefs
%\listfigs   %(if necessary)
\bye